\documentstyle[epsfig]{aipproc2}
\newcommand{\be}{\begin{equation}}
\newcommand{\ee}{\end{equation}}
\newcommand{\ba}{\begin{array}}
\newcommand{\ea}{\end{array}}
\begin{document}

\title{Wigner energy, odd-even mass staggering and the time-odd 
mean-fields\footnote{Plenary talk presented at Nuclear Structure'98, 
Gatlinburg, Tenn., USA, August 10-15, 1998}}

\author{Wojciech Satu{\l}a}

\address{Institute of Theoretical Physics, University of Warsaw, 
             ul. Ho\.za 69, 00-681 Warszawa, Poland \\
        Joint Institute for Heavy Ion Research, Oak Ridge, TN 
             37831-6374, USA \\
        Department of Physics, University of Tennessee, Knoxville, 
             TN 37996, USA}


\maketitle 

\begin{abstract}
Various properties of single-particle Hartree-Fock ground-state solutions
in  $N\sim Z$ nuclei are investigated. The emphasis is 
on a role of single-particle mean-field in 
odd-even mass staggering. It is shown that, unlike in 
traditional scenario originating from the Fermi gas or macroscopic
models, the symmetry energy contribution 
to odd-even mass staggering is nearly 
cancelled by the contribution coming from the average level density.  
It allows to construct indicators probing both pairing as well as 
mean-field components to the odd-even mass staggering. 
The impact of the single-particle Hartree-Fock field on Wigner energy 
and residual $pn$ interaction in odd-odd nuclei is also discussed.

\end{abstract}

\section*{Introduction}

Mean-field is considered as a standard nuclear model
to describe medium mass and heavy nuclei. Indeed, 
various applications demonstrated ability of 
this relatively simple concept to describe wide range 
of nuclear phenomena. 
However, the accuracy of mean-field results or 
predictions is far from satisfactory when confronted
with the present day, high precision experimental 
data. The predictions in many cases
differ substantially for various 
parametrizations of the same effective interaction and the 
situation calls for systematic program to
optimize the forces. Recent work of Chabanat {\it et al.}~\cite{cha97}
can be considered as the first step   
towards finding comprehensive parametrization of the 
Skyrme force~\cite{sky56}. Indeed, the so 
called Lyon (SLy) forces developed in Ref.~\cite{cha97} are 
very widely used and are considered
to be one of the best among the Skyrme forces available nowadays.

The aim of this work is to search for certain generic 
features of the ground-state Skyrme-Hartree-Fock (SHF) solutions
which may help in better understanding of basic properties of
this force through the nature of self-consistent solutions.
First section discusses a role of single-particle SHF mean-field in
odd-even mass staggering (OES) which, in atomic nuclei, is usually
attributed to pairing~\cite{boh58,sat98}. 
  Proper understanding of mean-field and pairing contributions to OES
 is crucial to better understand and parametrize both channels.
The second section investigates the possibility to obtain an additional 
binding energy in the $N=Z$ nuclei within single-particle 
self-consistent SHF model 
or conventional, i.e. including only $T=1,|T_z|=1$ 
pairing correlations, self-consistent Hartree-Fock-Bogolyubov
model. Finally, the third section discusses properties of self-consistent
SHF solutions in time-odd channel.

\section*{Odd-even mass staggering}

\begin{figure}[t!] 
\centerline{\epsfig{file=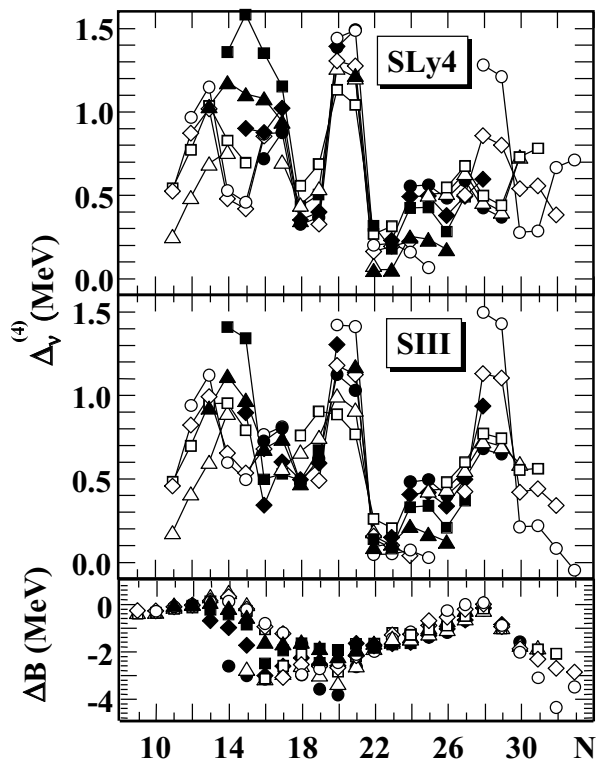,height=5.0in}} 
\caption{The OES $\Delta_\nu^{(4)}$ as a function of neutron number
         $N$ computed from theoretical masses
          calculated using SHF theory with SLy4 (upper part) and SIII 
          (middle part) Skyrme interaction. Note that the pattern of
          OES is almost independent
          on the interaction. The lowest panel shows the difference 
          of nuclear masses calculated in both cases. 
          Unlike the OES (local mass correlations) calculated masses
          are strongly interaction dependent.}
\vspace*{10pt}
\label{fig1}
\end{figure}

The odd-even mass staggering in atomic nuclei is usually attributed
to the presence of pairing correlations~\cite{boh58}. 
Indeed, in the simplest scenario based on standard BCS theory of 
superfluidity~\cite{bcs}, the ground state energies of three adjacent
isotopes (isotones) can be approximately interrelated by a simple 
readjustment of their chemical potentials $\lambda$:
\be 
B(N),\quad   B(N\pm 1)\approx B(N) + E_k \pm \lambda ,
\ee
where $E_k\approx\Delta $ denotes energy of the lowest quasiparticle 
in odd-$N$ nucleus. Consequently, the quantity 
\be\label{delta3}
\Delta^{(3)}_\nu (N)= 
{\pi_N\over 2}( B(N-1) -2 B(N) + B(N+1)  ) \approx E_k 
\approx \Delta ,
\ee
is often interpreted as a measure of empirical pairing gap, 
$\Delta$~\cite{rin80}. In Eq.~(\ref{delta3}), $\pi_N=(-1)^N$ is the
number parity and $B(N)$ is the (negative) binding energy of the 
system with $N$ particles. It has been noticed long time ago
based on the Fermi gas model that the quantity~(\ref{delta3})
contains strong contribution from the symmetry energy~\cite{bohrII}.
Because $\Delta^{(3)}_\nu (N)$ is a finite step approximation of 
derivative of the second order:
\be\label{deriv}
\Delta^{(3)}_\nu (N)\approx {\partial^2 B(N) \over \partial N^2},
\ee 
the symmetry energy [$\sim a_I(A)(N-Z)^2$] will indeed strongly 
contribute to OES according to this criterion.
However,
this contribution is independent on nucleon-number parity and,
therefore, the symmetry energy effect  can be 
removed simply by replacing the three-point indicator~(\ref{delta3})
by the  higher order expression like the one containing 
four masses~\cite{nil61}:
\begin{eqnarray}
\Delta^{(4)}(N) & \equiv &
{\pi_N\over 4}  \left[3B(N-1)-3 B(N)-B(N-2)\right. \nonumber \\
&+&\left.B(N+1)\right]
=\frac{1}{2} [\Delta^{(3)}(N) + \Delta^{(3)}(N-1)]  \label{delta4}
.
\end{eqnarray}
Consequently, in the traditional picture originating from the Fermi gas 
or liquid drop (macroscopic) models the mean-field contribution
to the odd-even mass staggering is predominantly due to 
symmetry energy and can essentially be removed by
means of high-order indicators like~(\ref{delta4}), see~\cite{jen84} and 
refs. quoted therein.

\begin{figure}[t!] 
\centerline{\epsfig{file=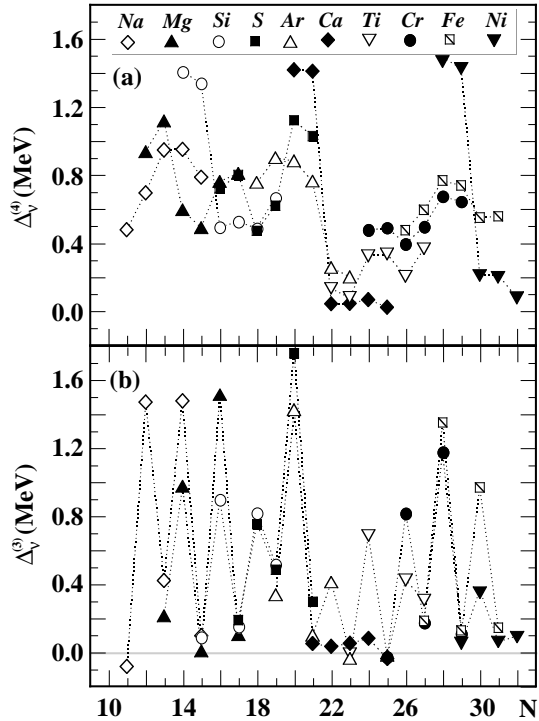,height=4.0in}} 
\caption{ The OES $\Delta_\nu^{(4)}$ (upper part) and $\Delta_\nu^{(3)}$
          (lower part) computed from theoretical masses
          calculated using SHF theory with SIII Skyrme interaction.
          Note, that rather complicated pattern of OES $\Delta_\nu^{(4)}$
          simplifies for $\Delta_\nu^{(3)}$. The $\Delta_\nu^{(3)}$
          shows clear alternating pattern  with large 
          values for even-$N$
          and small for odd-$N$. Taken from~\protect\cite{sat98}.}
\vspace*{10pt}
\label{fig2}
\end{figure}

  Contrary to the above so far commonly accepted scenario, the 
indicator~(\ref{delta4}) generates sizable OES when applied  to the 
mass table calculated using single-particle Skyrme-Hartree-Fock model,
see~\cite{sat98} and Fig.~\ref{fig1}.
These mass calculations has been performed for 
nuclei with  9$\leq$$Z$$\leq$28 and
$-2$$\leq$$N$$-Z$$\leq$6, using HFODD code (v1.75) of~\cite{dob97} 
and two parametrizations of the Skyrme force: SIII \cite{bei75}
 and SLy4~\cite{cha97}. 
The pattern of OES, which reaches from 30\% to 50\% of  experimental 
value, appears to be generic feature of the SHF solutions i.e. is almost 
independent on parametrization although calculated masses differ 
substantially for the two parametrizations employed, see Fig.~\ref{fig1}.

A rather complicated pattern of OES in Fig.~\ref{fig1}
simplifies when criterion $\Delta^{(3)}$ is used instead of $\Delta^{(4)}$,
see Fig.~\ref{fig2}. Indeed, values of $\Delta^{(3)}(N)$ are large for even-$N$
and small (close to zero) for odd-$N$. Furthermore, it appears possible
to explain this alternating $\Delta^{(3)}(N)$ 
behavior obtained directly in fully microscopic SHF calculations
by using simple arguments based on Strutinsky
energy theorem~\cite{str74}. According to this theorem the results of 
self-consistent calculations can be well approximated by the 
microscopic-macroscopic shell-correction method i.e.
the total binding energy can be written as
$B=E_{\rm sp} - \tilde{E}_{\rm sp} +E_{\rm macro}$,
where
\begin{equation}\label{sps}
E_{\rm sp} = {\textstyle{\sum_{k=1}^{A}e_k}}
\end{equation}
is the shell-model energy (sum of single-particle energies of occupied states),
$\tilde{E}_{\rm sp}$ is the Strutinsky-averaged shell-model energy,
and $E_{\rm macro}$ stands for the macroscopic
liquid-drop energy.

\begin{figure}[t!] 
\centerline{\epsfig{file=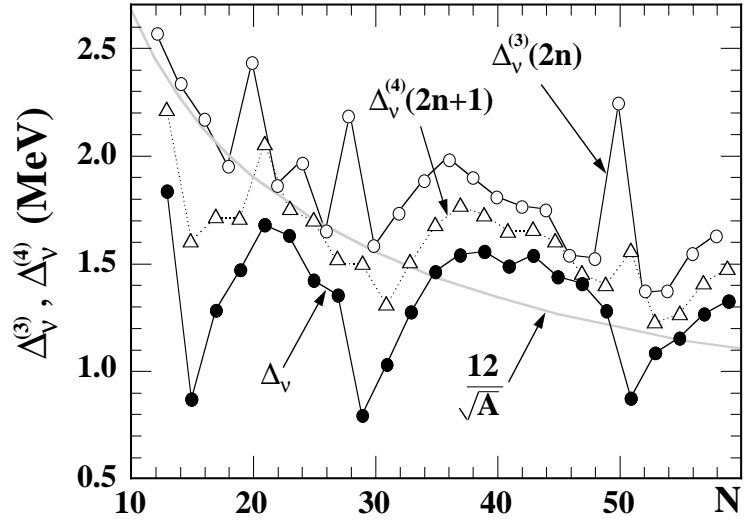,height=3.0in}}
\caption{  The empirical values of 
          $\Delta_\nu^{(3)}(N=2n+1)$[$\equiv \Delta_\nu$], 
          $\Delta_\nu^{(3)}(N=2n)$,
          and $\Delta_\nu^{(4)}(N=2n+1)$ as a function of neutron number.
          Each point represents average mean over
          several even-$Z$ isotones. Note, that only 
          $\Delta_\nu^{(3)}(N=2n+1)$ shows characteristic quenching at 
          (semi)magic gaps at $N=14,28,50$ (no effect at $N=20$ is seen) 
          as anticipated for pairing. Thick gray line shows the
          average trend $\Delta = 12/\protect\sqrt{A}$\,MeV. Taken from 
          Ref.~\protect\cite{sat98}.}
\vspace*{10pt}
\label{fig3}
\end{figure}

 The contribution to OES from shell-model energy is:
\begin{equation}\label{eeff}
\delta e\equiv \Delta^{(3)}_{\rm sp}(N) \approx
 {1\over 4} (1+\pi_N)(e_{n+1}-e_n) =
\left\{ \begin{array}{rrl} 0 & \mbox{if} & N=2n+1 \\
             (e_{n+1}-e_n) & \mbox{if} & N=2n   \end{array} \right.
\end{equation}
This single-particle mechanism behind OES was first
noticed in Ref.~\cite{bohrI} and applied to OES in metal 
clusters~\cite{cle85,man94,yan95}. The contribution coming from 
smooth Strutinsky energy can be expressed through average level
density at the Fermi energy $g(\lambda)$:
\begin{equation}\label{der}
\Delta^{(3)}(N) \approx
{\pi_N\over 2g(\lambda)}.
\end{equation}
The average single-particle level density is  $g(\lambda)=3a/\pi^2(m^*/m)$
\cite{bohrI}. The empirical value of the level density parameter
 for light nuclei is $a$$\approx$$A$/8\,MeV what agrees well with the estimate
based on  realistic
potentials \cite{shl92}. A correction due to effective mass is
$m^*/m=0.76$(0.70) for SIII and SLy4 forces, respectively.
Consequently, the contribution to OES coming 
from smooth Strutinsky energy equals
$\delta\Delta^{(3)}\approx -18/A$\,MeV. 
Contribution from macroscopic energy is essentially due to symmetry energy
[${a_I\over 2A}(N-Z)^2$]. The empirical
value of the symmetry energy  strength  
in light nuclei is $a_I=38$\,MeV~\cite{sat98b}. Hence, the liquid-drop
contribution to OES is $\delta\Delta^{(3)}\approx 19/A$\,MeV 
what indeed nearly
cancels out the contribution coming from the smooth Strutinsky energy.

The alternating  pattern of $\Delta^{(3)}(N)$ allows to extract both
pairing and mean-field component of OES. Indeed, 
at odd-$N$ mean-field component to $\Delta^{(3)}(N)$ is small 
and its value can be associated mainly with pairing:
\begin{equation}\label{pairing}
\Delta_\nu(N)\equiv\Delta_\nu^{(3)}(N=2n+1),
\end{equation} 
while the differences: 
\begin{equation}\label{nilsson}
e_{n+1} - e_{n} =2\left[\Delta_\nu^{(3)}(N=2n) - \Delta_\nu^{(3)}(N=2n+1)
\right],
\end{equation}
provide information about {\it effective Nilsson\/} single-particle 
spectra [for more details concerning this aspect see discussion 
and Fig.~4 in Ref.~\cite{sat98}].

 Empirical values of 
$\Delta_\nu^{(3)}$ and $\Delta_\nu^{(4)}$
are shown in Fig.~\ref{fig3}. Note, that only values of 
$\Delta_\nu^{(3)}(N=2n+1)$ show oscillatory behavior reflecting expected
quenching of neutron pairing correlations at the magic (semi-magic) 
gaps at $N$=14,28, and 50. Rather surprisingly no quenching effect is
seen at $N$=20. For $\Delta_\nu^{(3)}(2n)$ and $\Delta_\nu^{(4)}$
the expected quenching of pairing correlations 
is counterbalanced by single-particle effects and no minima are seen 
in these cases. These empirical 
observations neatly support the above scenario.

\begin{figure}[t!] 
\centerline{\epsfig{file=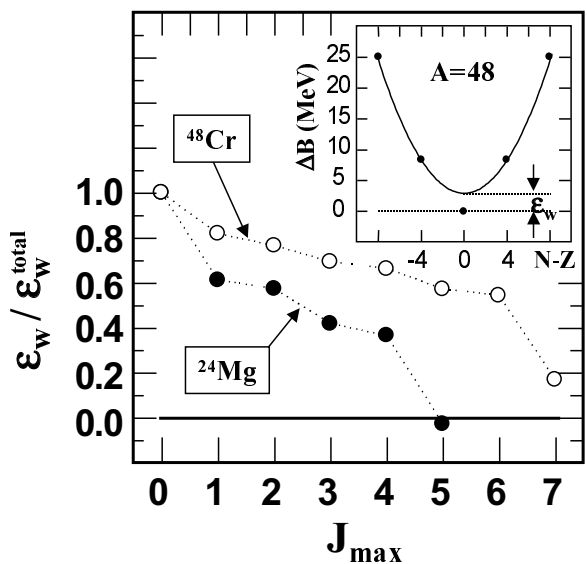,height=4.0in}}
\caption{ The contributions to the energy $\varepsilon_W$ 
          coming from isoscalar pairs of different angular momenta in
          $^{24}$Mg (solid circels) and $^{48}$Cr (open circels). 
          The energy $\varepsilon_W$ defines additional binding energy
          in $N=Z$ nucleus as a displacement of its binding energy
          from the average 
          parabolic $\sim (N-Z)^2$ behavior in the $N\ne Z$ isostones, 
          see inset. The details of this $0\hbar\omega$ shell-model 
          calculation can be found in Ref.~\protect\cite{sat97b}.}
\vspace*{10pt}
\label{fig4}
\end{figure}

  Similar interpretation of quantities (\ref{pairing}) and (\ref{nilsson})
can be drown also based on seniority, equidistant-level
[Richardson], and pairing-plus-quadrupole models~\cite{dob98}.
Let's consider here as an example a seniority (or pairing quasispin) model. 
The ground state energy in this model
[see e.g.~\cite{rin80} p. 222] equals:
\begin{equation}
B(N,s) = -{G\over 4}(N-s)(2\Omega-s-N+2)\quad\mbox{where}\quad \left\{
\begin{array}{cll} s=0 & \mbox{for} & N=2n \\
                   s=1 & \mbox{for} & N=2n+1 \end{array} \right. ,
\end{equation}
where $s$ is a seniority quantum number, $\Omega$ stands for 
degeneracy of the shell and $G$ denotes pairing strength.
After simple algebra one gets:
\begin{equation}\label{seniority}
\Delta^{(3)}(N)= \left\{
\begin{array}{lll} \displaystyle {1\over 2}G\Omega + {1\over 2}G & 
                                                   \mbox{for} & N=2n \\
                                                    &                \\
                   \displaystyle {1\over 2}G\Omega  & \mbox{for} & N=2n+1 
\end{array} \right. .
\end{equation}
Note that in accordance with (\ref{pairing}) indicator $\Delta^{(3)}(N=2n+1)$ 
probes only collective pairing energy 
while $\Delta^{(3)}(N=2n)$ contains also weak, $\delta e=G$ according
to (\ref{nilsson}), {\it mean-field\/} 
contribution. 

In the BCS approximation 
single-particle potential contributes both to $\Delta^{(3)}(N=2n)$
and $\Delta^{(3)}(N=2n+1)$. In this case one obtains Eq.~(\ref{seniority}) 
only in open shell regime where number of pairs is  
$n\sim\Omega/2 \gg 1$. In this regime, where BCS is considered
to be very good approximation, also 
$\Delta_{BCS}\approx G\Omega/2[\equiv\Delta^{(3)}(2n+1)_{exact}]$.

\section*{Wigner energy}

It has been known since the early work of~\cite{mye66} that 
macroscopic-microscopic
approaches systematically underbinds $N\approx Z$ nuclei. 
Similar situation holds in semi-classical Thomas-Fermi models
\cite{etfsi,mye97}
as well as in {\it spherical\/} Skyrme-Hartree-Fock calculations irrespectively
on parametrization of the Skyrme force~\cite{sat97b,sat98b}. 
This extra binding energy, which is characterized by $\sim |N-Z|$ type 
singularity at the $N=Z$ line, is dubbed Wigner energy
and is usually parametrized as:
\begin{equation}\label{wigner}
E_W=W(A)|N-Z| + d(A)\pi_{pn}\delta_{NZ},
\end{equation}
where $W(A)$ and $(A)$ are smooth functions of mass, and 
$\pi_{pn}=1$(0) for odd-odd(other) nuclei. 
Parametrization
(\ref{wigner}) can be justified based on simple arguments of
counting of $np$-pairs in identical spatial orbitals and on elementary
properties of $np$ interaction~\cite{mye77,jen84}.
This estimate gives also $W/d$=1. In Ref.~\cite{sat97b} (for more details
see~\cite{sat98b}) 
we constructed a family of indicators able to probe both 
$W(A)$ and $d(A)$. It appears that indeed $W(A)\approx d(A)$ 
provided that the energies of the lowest $T=0$ states are used 
in these indicators instead of 
masses of heavy $N=Z$ odd-odd nuclei. It provides independent 
empirical argument that Wigner energy is indeed predominantly related 
to $T=0$ $np$-interaction.

\begin{figure}[h!] 
\centerline{\epsfig{file=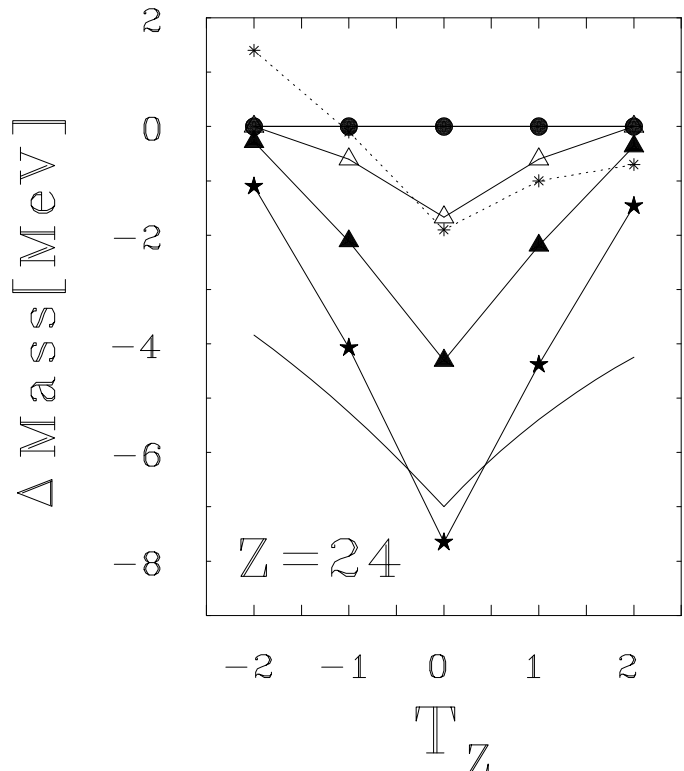,height=4.0in,width=3.5in}}
\caption{ The mass excess $\Delta M = B(x^T) - B(x^T=1)$ calculated using 
          generalized BCS plus Lipkin-Nogami theory as a function 
          of $T_z$. Different curves correspond to different values of
          $x^T$ being a ratio of the isoscalar to the isovector coupling
          constants. Dotted line marks the results of extended Thomas-Fermi
          calculations~\protect\cite{etfsi} while solid line without symbols 
          Wigner energy according to formula from 
          Ref.~\protect\cite{mye66}. Figure taken from 
          Ref.~\protect\cite{sat97a}.
}
\vspace*{10pt}
\label{fig5}
\end{figure}

Unlike mean-field, the state-of-the-art
shell-model calculations reproduce empirical Wigner energy 
very well~\cite{sat97b}. One can, therefore, use this model
to gain some knowledge about microscopic structure of the Wigner
energy. 
The shell-model indeed relates the Wigner energy
to $T=0$ interaction.
It has been demonstrated first by Brenner {et al.}~\cite{bre90}
in $sd$ shell. The detailed microscopic analysis in $sd$ and $fp$ 
shells~\cite{sat97b} revealed, however, rather  
complex structure of the Wigner energy in terms of nucleonic 
isoscalar pairs of different angular momenta. Although the largest
contributions come from pairs coupled to $J=1$ and  
$J=J_{max}=5$(7) for $sd$
and $fp$ shells, respectively, all matrix elements of intermediate 
angular momenta seem to be rather important, see Fig.~\ref{fig4}.

 The deficiencies of conventional mean-field which allows
only for $T=1$, $|T_z|$=1 pairing  can be 
remedied in generalized mean-field theory which takes into 
account also $np$-pairing correlations. It has been demonstrated
in Ref.~\cite{sat97a} that isoscalar $np$-pairing correlations 
can provide missing binding energy at $N=Z$ line and its closest
vicinity as shown in Fig.~\ref{fig5}. 
The model Hamiltonian employed in~\cite{sat97a} was very  
simple, based on similar pair counting mechanism  
like discussed above in connection with the Wigner energy 
parametrization.
For such a simple Hamiltonian generalized
BCS solutions do not mix $T=0$ and $T=1$ pair correlations
(see also~\cite{eng96}) and the energy gain is 
possible only when the isoscalar correlations are 
on the average stronger than the isovector.
 The latter cannot give
an extra binding energy (in even-even nuclei) as long as  isospin is 
(approximately) conserved.
Simple estimates of isoscalar and isovector pair correlations
in {\it simple nuclei\/} like $^{42}$Sc suggest indeed that   
isoscalar correlations are on the average stronger than isovector,
see discussion in~\cite{ros48,ana71}. However, connection between
empirical interactions deduced from the spectra of simple nuclei
and the mean-field effective pairing interaction is not obvious.  
Fast disappearance of isoparing with increasing
$T_z=(N-Z)/2$ shows up naturally in generalized BCS  model and
can be easily understood in terms of blocking of $np$-pairing
due to proton or neutron excess~\cite{sat97a,sat98c}.
One should mention here, that
recent generalized BCS calculations with
$G$-matrix interaction~\cite{goo98} in $fpg$ shell ($A\sim$80) do allow for 
mixing of $T=0$ and $T=1$ pair correlations similar to the shell-model 
solutions~\cite{eng96}.

\begin{figure}[t!] 
\centerline{\epsfig{file=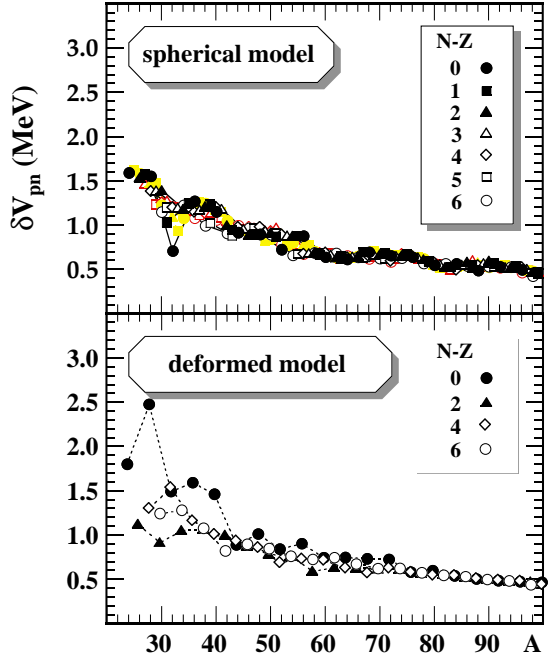,height=4.0in}}
\caption{  The values of $\delta V_{pn}$ (see Eq.~(\protect\ref{vpn})) 
          calculated
          using theoretical masses computed by means of {\it spherical\/} 
          SHF model with SLy4 Skyrme force and density dependent,
          surface active delta 
          interaction in pairing channel (upper part) and 
          theoretical masses computed 
          by Tajima {et al.}~
          \protect\cite{taj96}
          using {\it deformed\/} SHF-plus-BCS theory with 
          SIII Skyrme interaction (lower part). Figure taken 
          from~\protect\cite{sat98b}.} 
\vspace*{10pt}
\label{fig6}
\end{figure}

 The $np$-pairing is not necessarily the only contribution to the 
Wigner energy. In principle also  self-consistent mean-field itself 
can contribute to the extra binding in  $N=Z$ nuclei because of 
congruent nodal structure of a neutron and proton wave functions 
which can lead to stronger $np$-interaction as suggested by~\cite{mye97}.
The congruence energy can manifest itself only in fully microscopic 
models. Spherical, conventional Hartree-Fock-Bogolyubov
calculations with $T=1,|T_z|=1$ pairing only do not show any congruence 
effects as shown in Fig.~\ref{fig6}, see also
Refs.~\cite{sat97b,sat98b}. This result neither
depends on Skyrme interaction nor pairing force.
However, deformed SHF-plus-BCS calculations do show around
$\sim$30\% of the empirical Wigner energy.
Both spherical and deformed calculations are illustrated in Fig.~\ref{fig6} 
which shows values of the so called
double-difference indicator~\cite{zha89}
\begin{equation}\label{vpn}
\delta V_{np}= [B(N,Z)-B(N-2,Z)-B(N,Z-2)+B(N-2,Z-2)]/4\approx 
{\partial^2 B(N,Z)\over \partial N\partial Z}.
\end{equation}
This quantity is particularly convenient to probe 
Wigner energy~\cite{sat97b}.
In deformed calculations  (only even-even nuclei) masses computed by Tajima 
{\it et al.}~\cite{taj96} were used to calculate $\delta V_{np}$.
Very similar effect (about 30\% of experimental value) was obtained
also in our single-particle deformed SHF 
calculations~\cite{sat98b}. In these calculations
also odd-odd nuclei has been included. Rather surprisingly, an enhancement 
of $\delta V_{np}$
in odd-odd $N=Z$ nuclei comes entirely from the 
{\it time-odd isoscalar field}, see next section for greater detail.

\begin{figure}[t!] 
\centerline{\epsfig{file=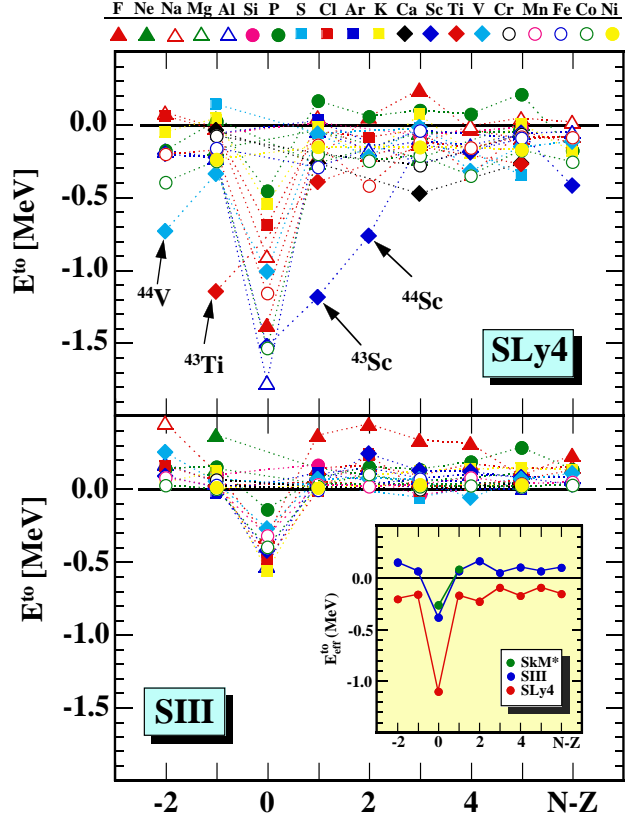,height=4.7in,width=3.5in}}
\caption{  The effective energetical contribution to nuclear mass  
          due to time-odd mean-field as a function of $N-Z$. 
          The calculations were performed using SHF model with Sly4 
          (upper part) and SIII
          (lower part) Skyrme forces, respectively. An inset represents
          configuration independent contributions i.e. arithmetic averages
          over $T_z$=const nuclei. }
\vspace*{10pt}
\label{fig7}
\end{figure}

\section*{The time-odd mean-fields}

Investigation of odd and odd-odd
nuclei require the time-odd mean-fields to be systematically 
taken into account in the calculations.
The time-odd part of the energy density ${\cal H}^{odd}$
reads:
\begin{equation}\label{odd}
{\cal H}^{odd}_t= C^{s}_{t}{\mbox{\boldmath $s$}}^2_t+
 C^{\Delta s}_{t}{\mbox{\boldmath $s$}}_t 
  \cdot {\mbox{\boldmath $\Delta s_t$}} +
                 C^{T}_{t} {\mbox{\boldmath $s_t\cdot T_t$}} +
                 C^{j}_{t}{\mbox{\boldmath $j$}}^2_t
 C^{\nabla j}_{t} {\mbox{\boldmath $s_t\cdot (\nabla \times j_t)$}} 
\end{equation}
where {\boldmath $s, j, T$} are time-odd spin, momentum,
and vector kinetic densities, respectively, and subscript $t$[=0,1] denotes 
isospin~\cite{dob95,dob97}.

The effective contribution of the time-odd mean-fields 
to the total energy $E^{to}$ 
can be investigated by comparing full SHF 
calculations with the calculations restricted to time-even fields only. 
The result is shown in Fig.~\ref{fig7}. As seen from the figure everywhere
beyond $N=Z$ line $E^{to}$  is small 
but force dependent. 
Indeed, the time-odd fields are effectively attractive for SLy4 but repulsive 
for SIII forces, respectively. Also beyond $N=Z$ line simple additivity 
for the averages [see inset in Fig.~\ref{fig7}]: 
\begin{equation}\label{add}
\overline{E}^{to}(N-Z=2n)\approx 
\overline{E}^{to}(N-Z=2n-1) + 
\overline{E}^{to}(N-Z=2n+1),
\end{equation}
is rather well fulfilled, reflecting single-particle nature of the effect.
The most remarkable observation is an enhancement of the 
time-odd effects in $N=Z$ nuclei. The effect is essentially due to 
cancellation of
two large components in (\ref{odd}), namely a 
repulsive $C^{s}_{0}${\boldmath $s^2_0$} field is overbalanced by an 
attractive 
$C^{\Delta s}_{0}${\boldmath $s_0 \cdot \Delta s_0$}
field.
The effect reflects most likely spontaneous spin polarization at the nuclear 
surface but its nature and consequences 
are not yet fully recognized and are under study~\cite{sat98d}.
It is noteworthy, that
any observable consequences of this {\it spin collectivity\/} may help to 
establish certain 
constraints on time-odd coupling 
constants $C^{s}_{0}$ and $C^{\Delta s}_{0}$.
Note that $C^{s}_{t}$ and $C^{\Delta s}_{t}$
are the only {\it free\/} time-odd coupling constants for Skyrme force.
The remaining
coupling constants (\ref{odd}) are related to time-even coupling constants
because of local gauge invariance of the Skyrme force~\cite{eng75}.

It is interesting to observe also strong instabilities in time-odd fields 
for $N\ne Z$ $^{43}$Sc, $^{44}$Sc nuclei and in their isobaric analogs 
$^{43}$Ti, $^{44}$V. These instabilities show up only for SLy4 force and
only at the very bottom of the $f_{7/2}$ shell. This is, most likely, due
to the extended spatial dimensions of these nuclei and due to strong
attractiveness of $C^s_0$ component of the SLy4 force at low
densities where $C^s_0(\rho \rightarrow 0)= -207.8$\,MeV\,fm$^{3}$.
For most of the commonly used Skyrme forces 
$C^s_0(\rho \rightarrow 0)>0$, see table I in Ref.~\cite{dob95}.

Let's finally consider briefly an indicator
\begin{eqnarray}\label{pn}
\epsilon_{pn}=(-1)^{(A+1)}[-B(N-1,Z-1)&+&2B(N,Z-1)-B(N+1,Z-1)+ \nonumber \\
                           2B(N-1,Z)&-&4B(N,Z)+2B(N+1,Z) \nonumber \\
                         -B(N-1,Z+1)&+&2B(N,Z+1)-B(N+1,Z+1)]/4,
\end{eqnarray}
which probes residual {\it pn}-interaction in odd-odd nuclei, see~\cite{jen84}.
The data indicate, that for light $N\sim Z$ nuclei 
$\epsilon_{pn}$ is weekly dependent on nuclear mass and
$\epsilon_{pn}\approx 500$\,keV. Only in the closest vicinity of 
$N=Z$ line $\epsilon_{pn}$ is enhanced due to the Wigner energy, see
\cite{sat97b,sat98b}. Applying the indicator (\ref{pn}) to single-particle
SHF mass table 
gives {\it no effect\/} i.e. $\epsilon_{pn}\approx 0$ in odd-odd 
$N\ne Z$ nuclei. 
The cancellation of time-odd effects in (\ref{pn}) is a consequence
of the above mentioned, see Eq.~(\ref{add}), energetical additivity 
of time-odd contributions in odd and odd-odd nuclei. 
Only in $N = Z$ odd-odd cases $\epsilon_{pn}\ne 0$ as a result of
an enhancement of the time-odd fields in these nuclei.
It seems therefore justified to conclude that the residual neutron-proton 
interaction in odd-odd nuclei as seen through the indicator
(\ref{pn}) goes entirely beyond the mean-field.

\section*{Summary}

Selected properties of single-particle self-consistent Skyrme 
mean-field have been analyzed. It has been shown by direct microscopic 
calculations that the mean-field component to OES according to the criterion
(\ref{delta3}) shows strong nucleon number parity $\pi_{N(Z)}$
dependence with $\Delta_{\nu(\pi)}^{(3)}$ being large (small) for 
$\pi_{N(Z)}=+1$(-1), respectively. 
This alternating pattern is due to ({\it i\/}) cancellation of the 
contributions from symmetry energy and average level density, and 
({\it ii\/}) due to nuclear Jahn-Teller effect. It 
allows to interpret $\Delta_{\nu(\pi)}^{(3)}(N=2n+1)$
directly as due to (almost) pure pairing and construct higher 
order indicator (\ref{nilsson}) to probe also the mean-field component
to OES. This scenario is entirely different
than the commonly accepted scenario derived from the
 Fermi gas model where mean-field component to OES
is due to symmetry energy and thus is independent on the number parity.

It has been demonstrated that {\it spherical\/} HFB calculations 
including only $T=1$,$|T_z|=1$ pairing generate no congruence
energy i.e. $W(A)\approx 0$. It appears also, that 
{\it deformed\/} single-particle SHF (or SHF-plus-BCS) mean-field  
can generate only relatively small fraction $\sim$30\% of the 
empirical Wigner energy strength $W(A)$  in even-even $N=Z$ nuclei. 
Therefore, most of the Wigner energy strength in $N=Z$ nuclei  
seems to be beyond the Skyrme mean-field and, most likely, is due to 
neutron-proton pairing.

One of the most spectacular observations is strong enhancement 
of the time-odd effects in $N=Z$ nuclei. Closer examination shows that
this effect depends essentially on the first two
components (more precisely is due to their isoscalar parts) in (\ref{odd}). 
Because $C^{s}_{0}$ and
 $C^{\Delta s}_{0}$ are
independent (free) time-odd coupling constants for Skyrme force 
any observable consequences of this effect can help to establish empirical
constraints for their values. 
In contrast to the $N=Z$ nuclei, in $N\ne Z$ nuclei effective  contributions 
to the total binding energy coming from the time-odd fields
reflect simple additivity pattern between odd and odd-odd nuclei typical for a
single-particle model. As a consequence, binding energy between
valence proton and neutron in odd-odd nuclei meassured by means of the
indicator (\ref{pn}) is {\it zero\/}. In other words $\epsilon_{pn}$ 
is entirely beyond SHF field.

\section*{Acknowledgments}

The material presented here was obtained in collaboration with D.J. Dean, 
J. Dobaczewski, W. Nazarewicz, and R. Wyss.
This research was supported in part by
the U.S. Department of Energy
under Contract Nos. DE-FG02-96ER40963 (University of Tennessee),
DE-FG05-87ER40361 (Joint Institute for Heavy Ion Research),
DE-AC05-96OR22464 with Lockheed Martin Energy Research Corp. (Oak
Ridge National Laboratory), and  by the Polish Committee for
Scientific Research (KBN) under Contract No.~2~P03B~040~14.



\begin{references}

\bibitem{cha97} E. Chabanat E., Bonche P., Haensel P., Meyer J., and 
Schaeffer R., {\it Nucl. Phys} {\bf A627}, 710 (1997).

\bibitem{sky56} Skyrme T.H.R., {\it Phil. Mag.} {\bf 1}, 1043 (1956);
{\it Nucl. Phys.} {\bf 9}, 615 (1959).

\bibitem{boh58} Bohr A., Mottelson B.R., and Pines D., {\it Phys. Rev.} 
{\bf 110}, 936 (1958).

\bibitem{sat98}
Satu{\l}a W., Dobaczewski J., and Nazarewicz W., 
{\it Phys. Rev. Lett.} in print, and nucl-th/9804060 

\bibitem{bcs} Bardeen J., Cooper L.N., and Schrieffer J.R., 
{\it Phys. Rev.} {\bf 110}, 1175 (1957) 

\bibitem{rin80} Ring P., and Schuck P., {The Nuclear Many-Body Problem},
Berlin: Springer-Verlag, (1980)

\bibitem{bohrII}
{Bohr A., and Mottelson B.R., {\it Nuclear Structure}, vol. 2, 
New York: W.A. Benjamin, 1975}.

\bibitem{nil61} Nilsson S.G., and O. Prior, {\it Mat. Fys. Medd. Dan. 
Vid. Selsk.} {\bf 32}, No. 16 (1961).

\bibitem{jen84} Jensen A.S., Hansen P.G., and Jonson B., 
{\it Nucl. Phys.} {\bf A431}, 393 (1984).

\bibitem{dob97}
Dobaczewski J., and Dudek J., {\it Comp. Phys. Commun.}
{\bf 102}, 166, 183 (1997), and to be published.

\bibitem{bei75} Beiner M., Flocard H., Nguyen Van Giai, and Quentin P.,
{\it Nucl. Phys.} {\bf A238}, 29 (1975).

\bibitem{str74}
Strutinsky V.M., {\it Nucl. Phys.} {\bf A218}, 169 (1974).

\bibitem{bohrI}
Bohr A. and Mottelson B.R., {\it Nuclear Structure}, vol. 1, 
New York: W.A. Benjamin,  1969.

\bibitem{cle85} Clemenger K., {\it Phys. Rev.} {\bf B32}, 1359 (1985).

\bibitem{man94} 
Manninen M., Mansikka-aho J., Nishioka H., and  Takahashi Y.,
{\it Z. Phys.} {\bf D31}, 259 (1994).

\bibitem{yan95} 
Yannouleas C., and Landman U., {\it Phys. Rev.} {\bf B51}, 1902 (1995).

\bibitem{shl92}
{Shlomo S., {\it Nucl. Phys.} {\bf A539}, 17 (1992)}.

\bibitem{dob98}  Dobaczewski J., Nazarewicz W., and Satu{\l}a W., to be 
published.

\bibitem{mye66} Myers W., and Swiatecki W., {\it Nucl. Phys.} {\bf A81}, 1 
(1966).

\bibitem{etfsi} Abboussir Y., Pearson J.M., Dutta A.K., and Tondeur F.,
{\it Atomic Data and Nuclear Data Tables} {\bf 61}, 127 (1995)

\bibitem{mye97} Myers W., and Swiatecki W., {\it Nucl. Phys.} {\bf A612},
249 (1997).

\bibitem{sat97b} Satu{\l}a W., Dean D.J., Gary J., Mizutori S., and 
Nazarewicz W., {\it Phys. Lett.} {\bf B407}, 103 (1997).

\bibitem{sat98b} Satu{\l}a W., Dean D.J., Dobaczewski J., Garrett J., 
and Nazarewicz W., to be published.

\bibitem{mye77} Myers W., {\it Droplet Model of Atomic Nuclei}, 
New York: Plenum, 1977.

\bibitem{bre90} Brenner D.S, Wesselborg C., Casten R.F., Warner D.D.,
and Zhang J.-Y., {\it Phys. Lett.} {\bf B243}, 1 (1990). 

\bibitem{sat97a} Satu{\l}a W., and Wyss R.A., {\it Phys. Lett.} {\bf B393},
1 (1997).

\bibitem{ros48} Rosenfeld L., {\it Nuclear Forces}, 
Amsterdam: North-Holland, 1948. 

\bibitem{ana71} N. Anantaraman N., and Schiffer J.P., {\it Phys. Lett.} 
{\bf B37}, 229 (1971), and Schiffer J.P., {\it Ann. Phys.} {\bf 66}, 78 (1971).

\bibitem{sat98c} Satu{\l}a W., and Wyss R., to be published.

\bibitem{taj96} Tajima N., Takahara S., and Onishi N., {\it Nucl. Phys.}
{\bf A603}, 23 (1996).

\bibitem{goo98} Goodman A.L, {\it these proceedings}. 

\bibitem{eng96} Engel J., Langanke K., and Vogel P., {\it Phys. Lett.}
{\bf B389}, 211 (1996).

\bibitem{dob95} Dobaczewski J., and Dudek J., {\it Phys. Rev.} {\bf C52},
1827 (1995).

\bibitem{eng75} Engel Y.M., Brink D.M., Goeke K., Krieger S.J., and
D. Vautherin, {\it Nucl. Phys.} {\bf A249}, 215 (1975).

\bibitem{zha89} Zhang J.-Y., Casten R.F., and Brenner D.S., {\it Phys. Lett.}
{\bf B227}, 1 (1989). 

\bibitem{sat98d} Satu{\l}a W., Dobaczewski J., and Nazarewicz W., 
to be published.




\end{references}
\end{document}